\documentclass[aps,prb,reprint,twocolumn,showpacs,floatfix,raggedbottom]{revtex4-2}

\usepackage[T1]{fontenc}
\usepackage[utf8]{inputenc}
\usepackage{amsmath,amssymb,bm}
\usepackage{graphicx}
\usepackage{xcolor}
\usepackage[colorlinks=true,linkcolor=blue,citecolor=blue,urlcolor=blue]{hyperref}

\begin{document}

\title{Algebraic flat bands and even central-band Chern numbers in a seven-orbital hopping model}
\author{Weizhuo Xu}
\author{Jiachen Qian}
\author{Shuaihua Zhang}
\author{Yanling Wu}
\author{Jun Li}
\email{ljcj007@ysu.edu.cn}
\affiliation{State Key Laboratory of Metastable Materials Science and Technology,
Hebei Key Laboratory of Microstructural Material Physics, School of Science,
Yanshan University, Qinhuangdao, 066004, China}
\date{\today}

\begin{abstract}
Turning a flat-band parent into an isolated Chern band requires supervise
of destructive interference, hopping phases, and the cell translations
of the bonds. We study this connection in a seven-orbital 2$\times$2 supercell
honeycomb-vacancy lattice, combining
analytic reductions with Bloch spectra, lattice Chern invariants, and
quantum geometry. The parent lattice has three exact flat branches, a smooth
periodic zero mode, and a compact localized state. A uniform gap bound
shows that faint intracell-loop hopping preserves two topologically
trivial insulating gaps. The intracell hopping graph also admits a
staggered chiral coloring: reversing site amplitudes according to this
coloring links opposite energies at momenta separated by half a
reciprocal period. This constraint restricts the isolated central band
to even Chern numbers and forbids a zero-energy insulating filling at
every loop phase. Purely imaginary loop hopping admits a real gauge,
whereas other phases allow paired band contacts connecting central bands
with \(C_3=2\) and \(C_3=4\). On a specified strong-loop cut, third-order masses open the direct gaps;
these gaps shrink faster than the bandwidth as the coupling grows.
Changing the bond connectivity to an oriented honeycomb next-nearest-neighbor
extension generates a narrower Chern band with nonuniform quantum geometry.
These results distinguish individual-band topology from
occupied-group topology and reveal how bond connectivity and cell
translations determine which topological bands can emerge from an
exactly flat parent.
\end{abstract}

\maketitle

\section{Introduction}

When a band becomes flat, the kinetic-energy cost of redistributing particles
within it is suppressed. Interactions can then compete with the remaining
single-particle scales, as illustrated by correlated superconductivity and
Chern insulators in moir\'e graphene~\cite{Cao2018,Chen2020}.
Rigorous Hubbard-model results connect sublattice imbalance and engineered
band degeneracies to ground-state spin, ferrimagnetism, and ferromagnetism
at particular fillings~\cite{Lieb1989,Tasaki1992}.
These examples make the wavefunctions as important as the bandwidth:
they decide how particles overlap, how interactions act within the band,
and whether the band carries a topological invariant.

Destructive interference supplies a local route to flatness. Amplitudes
leaving a finite cluster can cancel, producing a compact eigenstate whose
translations generate a flat branch~\cite{Sutherland1986,Flach2014}.
This cancellation principle also underlies constructive flat-band generators
in one dimension~\cite{Maimaiti2017}.
Whether these states span the band, and whether the band touches its
neighbors, depends on real-space connectivity~\cite{Bergman2008}
and the regularity of its Bloch wavefunctions~\cite{Rhim2019,Rhim2020}.
In a bipartite lattice with $N_A>N_B$, the rectangular coupling block
necessarily has a kernel. This preserves a zero mode when onsite and
same-sublattice terms are absent~\cite{Ramachandran2017}.
Chern topology imposes a different requirement on wavefunctions across
the whole Brillouin zone. Strictly finite-range periodic hopping cannot
support a band that is simultaneously isolated, exactly flat, and
Chern nontrivial~\cite{Chen2014NoGo}.
The global character of this obstruction is also apparent from Wannier
functions: a nonzero-Chern band cannot be spanned by compactly supported
Wannier functions, even if the spanning set is overcomplete~\cite{Read2017}.
A topological extension of such a
flat-band parent must thus allow dispersion and control the gaps
to neighboring bands.

Nearly flat Chern bands~\cite{Sun2011,Tang2011} provide lattice settings for
interacting fractional Hall states~\cite{Neupert2011,Wang2011}.
Fractional Chern insulators have been observed at finite magnetic field
in graphene heterostructures and twisted bilayer graphene~\cite{Spanton2018,Xie2021}.
Optical signatures~\cite{Cai2023} and fractionally quantized
transport~\cite{Park2023} in twisted MoTe$_2$ extend this motivation
to zero-field settings. The relevant comparison relates bandwidth,
interaction strength, and interband separation together.
Wavefunction geometry supplies another part of this comparison:
the quantum metric contributes to flat-band superfluid
response~\cite{Peotta2015,Julku2016}, while the variation of curvature
and metric constrains fractional-band design~\cite{Roy2014}.
For constant Berry curvature in the long-wavelength limit, the projected
density operators follow the same algebra as in the lowest Landau
level~\cite{Parameswaran2012}, giving curvature uniformity a physical role.
A narrow band is thus a starting point for a many-body problem, whose
ground state also depends on its wavefunctions and interactions.

Complex hopping can generate Chern topology through the Haldane
mechanism~\cite{Haldane1988}; spin-orbit coupling provides related routes
on Lieb, perovskite, and kagome lattices~\cite{Weeks2010,Guo2009}.
Kagome Chern magnetism connects band topology to magnetic
response~\cite{Yin2020}, and bands with higher Chern numbers allow
extra correlated structures~\cite{Liu2012,Barkeshli2012,Bergholtz2013}.
For a spinless hopping model, the phases of closed hopping products and
the surviving symmetries determine whether complex amplitudes actually
break time reversal. A further distinction concerns occupation: a direct
gap separates a band from its neighbors at each momentum, whereas an
indirect gap allows one Fermi energy to lie in a gap throughout the zone.
The latter is required for an insulating filling and its quantized
occupied-group Hall response~\cite{Thouless1982,Xiao2010}.

Artificial lattices provide ways to control these ingredients~\cite{Leykam2018}.
Electronic and atomic implementations motivate tunable hopping
networks~\cite{Polini2013,Drost2017,Jotzu2014}, and photonic lattices
allow topological propagation to be studied in a different wave
setting~\cite{Rechtsman2013}.
Localized flat-band states have also been prepared and observed in
photonic Lieb lattices~\cite{Mukherjee2015}.
Here we remove one site from a honeycomb supercell to obtain a seven-orbital
parent with exact flat bands. We treat it as a spinless hopping network
that isolates the effects of connectivity, and follow how same-sublattice
hopping changes its flat bands.
An intracell loop leads to exact symmetry constraints, even Chern numbers,
and paired band contacts at topological transitions. Its strong-coupling limit exposes a further issue:
the leading dispersion can miss the small mass that isolates a topological
band. A honeycomb next-nearest-neighbor extension alters the bond
connectivity and provides complementary Chern bands, quantum geometry,
and boundary spectra. Together, these results identify which hopping
choices control band topology, spectral isolation, and insulating filling.

\section{Model and methods}

\begin{figure*}[t]
	\centering
	\includegraphics[width=\textwidth]{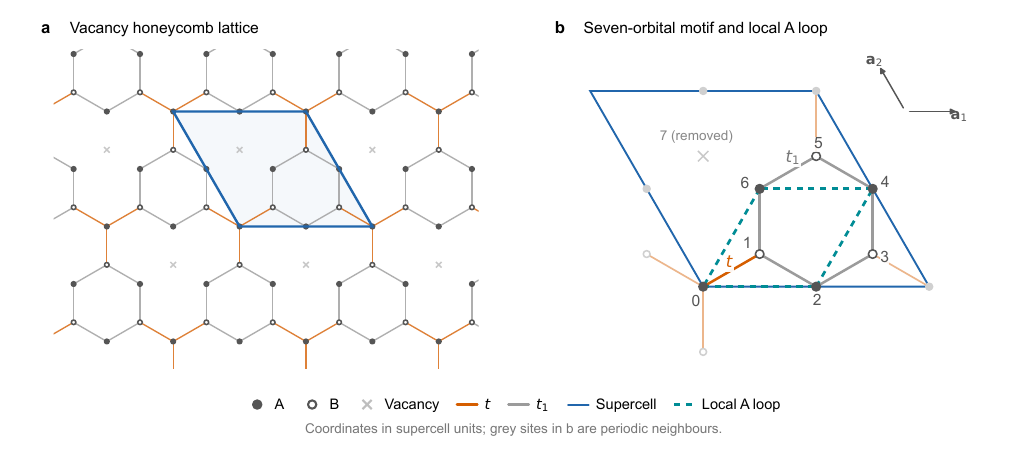}
	\caption{Honeycomb-vacancy geometry.
		(a) Repeated nearest-neighbor network with one supercell highlighted.
		Filled/open circles denote $A$/$B$ sites; crosses mark the vacancy.
		Orange/gray bonds carry $t$/$t_1$.
		(b) Orbital $0$ lies outside the ring $1,\ldots,6$; gray sites are periodic images.
		The dashed physical $A$ rhombus uses $\mathbf r_4+(1,0)$ and Eq.~\eqref{eq:physicalAA}.
		The intracell loop and NNN terms define different hopping extensions.}
	\label{fig:lattice}
\end{figure*}
\subsection{Seven-orbital vacancy lattice and hopping extensions}
\label{sec:model}
\label{sec:geometry}

The spinless, noninteracting parent is a honeycomb vacancy supercell with
$A=\{0,2,4,6\}$ and $B=\{1,3,5\}$ (Fig.~\ref{fig:lattice}).
The supercell basis is $\mathbf a_1=(1,0)$,
$\mathbf a_2=(-1/2,\sqrt3/2)$; the parent primitive vectors are
$\mathbf a_1/2,\mathbf a_2/2$.
Fractional orbital representatives are
\begin{equation}
\begin{aligned}
\mathbf r_0&=(0,0),&\mathbf r_1&=(1/3,1/6),\\
\mathbf r_2&=(1/2,0),&\mathbf r_3&=(5/6,1/6),\\
\mathbf r_4&=(0,1/2),&\mathbf r_5&=(5/6,2/3),\\
\mathbf r_6&=(1/2,1/2),&\mathbf r_7&=(1/3,2/3).
\end{aligned}
\label{eq:positions}
\end{equation}
Removing $B$ orbital $7$ remains nine nearest-neighbor bonds of length
$\sqrt3/6$. Their common embedded hopping matrix is
$[H_0(\boldsymbol\xi)]_{ij}=-t_{ij}e^{i\boldsymbol\xi\cdot\boldsymbol\delta_{ij}}$,
with directed entries listed in Table~\ref{tab:bonds}. The Hamiltonian includes their Hermitian reverses.
Subtracting the orbital offsets gives the integer cell translations
$\mathbf R_{ij}=\boldsymbol\delta_{ij}+\mathbf r_i-\mathbf r_j$.
For physical reciprocal coordinates $k_\mu=\mathbf k\cdot\mathbf a_\mu$,
$H_{\rm hc}^{\rm per}(\mathbf k)=D_gH_0(\mathbf k)D_g^\dagger$ is periodic
on $[-\pi,\pi)^2$, where $(D_g)_{ii}=e^{i\mathbf k\cdot\mathbf r_i}$.
The supercell path uses $K=(2\pi/3,2\pi/3)$ and $M=(\pi,0)$.

\begin{table}[t]
\caption{Directed parent bonds in the embedded triangular coordinates.
Hermitian reverse entries are included in the Hamiltonian.}
\label{tab:bonds}
\begin{ruledtabular}
\begin{tabular}{cccc}
$(i,j)$&$t_{ij}$&$\delta_x$&$\delta_y$\\
$(0,1)$&$t$&$1/3$&$1/6$\\
$(0,3)$&$t$&$-1/6$&$1/6$\\
$(0,5)$&$t$&$-1/6$&$-1/3$\\
$(1,2)$&$t_1$&$1/6$&$-1/6$\\
$(1,6)$&$t_1$&$1/6$&$1/3$\\
$(2,3)$&$t_1$&$1/3$&$1/6$\\
$(3,4)$&$t_1$&$1/6$&$1/3$\\
$(4,5)$&$t_1$&$-1/6$&$1/6$\\
$(5,6)$&$t_1$&$-1/3$&$-1/6$
\end{tabular}
\end{ruledtabular}
\end{table}

The intracell extension adds momentum-independent entries $H_{02}=H_{24}=H_{46}=H_{60}=z$,
with $z=t_Ae^{i\phi}$, to the embedded matrix.
Its periodic coordinates are $\mathbf q=\boldsymbol\kappa/2$.
Define $D(\boldsymbol\kappa)=\operatorname{diag}
(1,e^{-i\alpha},1,e^{-i\beta},1,e^{-i\gamma},1)$,
where $\alpha=\kappa_x/3+\kappa_y/6$,
$\beta=(-\kappa_x+\kappa_y)/6$, and $\gamma=-\kappa_x/6-\kappa_y/3$.
Then
\begin{equation}
\widetilde H(\mathbf q)=D^\dagger(2\mathbf q)
[H_0(2\mathbf q)+H_A]D(2\mathbf q)
\label{eq:periodic}
\end{equation}
has, in the basis $(0,2,4,6;1,3,5)$, the block form
\begin{equation}
\widetilde H(\mathbf q)=
\begin{pmatrix}A_\phi&M(\mathbf q)\\M^\dagger(\mathbf q)&0_{3\times3}\end{pmatrix},
\label{eq:full}
\end{equation}
where
\begin{equation}
M(\mathbf q)=-\begin{pmatrix}
t&t&t\\
t_1e^{-iq_x}&t_1e^{iq_x}&0\\
0&t_1e^{-iq_y}&t_1e^{iq_y}\\
t_1e^{-i(q_x+q_y)}&0&t_1e^{i(q_x+q_y)}
\end{pmatrix},
\label{eq:M}
\end{equation}
and
\begin{equation}
A_\phi=\begin{pmatrix}
0&z&0&z^*\\z^*&0&z&0\\0&z^*&0&z\\z&0&z^*&0
\end{pmatrix}.
\label{eq:Aphi}
\end{equation}
The complete intracell-loop torus is $\mathbf q\in[-\pi,\pi)^2$,
corresponding to $4\pi$ sewing periods in $\boldsymbol\kappa$.
Hopping translations fix this intracell integration domain, different from the physical supercell torus.

A loop following the spatial rhombus uses $\mathbf r_4+(1,0)$ and includes bond-displacement phases:
\begin{equation}
\begin{pmatrix}
(H_{\rm hc})_{02}\\(H_{\rm hc})_{24}\\
(H_{\rm hc})_{46}\\(H_{\rm hc})_{60}
\end{pmatrix}
=z\begin{pmatrix}
e^{ik_x/2}\\e^{i(k_x+k_y)/2}\\
e^{-ik_x/2}\\e^{-i(k_x+k_y)/2}
\end{pmatrix}.
\label{eq:physicalAA}
\end{equation}
Hermitian reverse hoppings are included.
Their canonical translations are $(0,0),(1,0),(-1,0),(0,0)$.
The loops share a nearest-neighbor parent but connect different cell images.
An orbital-gauge change cannot make the full hopping graphs equivalent.
The honeycomb NNN extension substitutes this rhombus by the oriented
hoppings in Eq.~\eqref{eq:NNN}, including $B$--$B$ terms.
The vacancy parent has threefold rotation and vertical mirrors,
whereas the selected four-edge loop breaks threefold rotation.

\subsection{Band isolation, topology, and geometry}
\label{sec:numerics}

Energies use $|t_1|$ units, with $t=t_1=1$ unless specified.
Hermitian eigenvalues are ordered $E_0\leq\cdots\leq E_6$; the central band is $n=3$.
Adjacent direct and indirect gaps are
\begin{equation}
\delta_n=\min_{\mathbf q}[E_{n+1}(\mathbf q)-E_n(\mathbf q)],
\label{eq:direct}
\end{equation}
\begin{equation}
\Delta_n=\min_{\mathbf q}E_{n+1}(\mathbf q)-\max_{\mathbf q}E_n(\mathbf q).
\label{eq:indirect}
\end{equation}
Positive direct gaps isolate subspaces at each momentum;
positive indirect gaps allow insulating filling across the zone.
For $w_3=\max E_3-\min E_3>0$ and both adjacent indirect gaps positive,
the flatness ratio is $f=\min(\Delta_2,\Delta_3)/w_3$.
Honeycomb quantities use $\mathbf k$ on its physical supercell torus.

Chern numbers use the Fukui--Hatsugai--Suzuki links~\cite{Fukui2005},
$U_\mu=\langle u(\mathbf q)|u(\mathbf q+\hat\mu)\rangle/
|\langle u(\mathbf q)|u(\mathbf q+\hat\mu)\rangle|$, with one mesh step $\hat\mu$:
\begin{equation}
\begin{aligned}
C=\frac{1}{2\pi}\sum_{\mathbf q}\operatorname{Arg}
\big[&U_x(\mathbf q)U_y(\mathbf q+\hat x)\\
&\times U_x^*(\mathbf q+\hat y)U_y^*(\mathbf q)\big].
\end{aligned}
\label{eq:FHS}
\end{equation}
For an isolated group, normalized determinant overlaps
$\det[V_r^\dagger(\mathbf q)V_r(\mathbf q+\hat\mu)]$
replace individual overlaps and track the whole subspace through internal band touchings;
$C_{\mathrm{low},3}$ and $C_{\mathrm{low},4}$ denote the lowest three and four bands.
We use the appropriate periodic gauge and meshes up to $192^2$,
then refine candidate extrema from multiple grid seeds; narrow strong-loop contacts require local resolution.
Extrema are numerical estimates, not certified comprehensive bounds.
Displayed results use three significant figures; calculations and source data retain full precision.

In the canonical periodic convention, the central-band quantum geometric tensor
describes how its eigenstate varies with momentum. Its real part measures
the local distance between quantum states~\cite{Provost1980}.
Analytic Hamiltonian derivatives give
\begin{equation}
Q_{\mu\nu}=\sum_{m\ne3}
\frac{\langle u_3|\partial_\mu H|u_m\rangle
\langle u_m|\partial_\nu H|u_3\rangle}{(E_3-E_m)^2},
\label{eq:QGT}
\end{equation}
with $\Omega_{xy}=2\operatorname{Im}Q_{xy}$ and $g_{\mu\nu}=\operatorname{Re}Q_{\mu\nu}$
for the orientation in Eq.~\eqref{eq:FHS}.
The triangular coordinates are not orthonormal, so honeycomb geometry uses Cartesian momenta
$p_x=k_x$, $p_y=(k_y+k_x/2)/(\sqrt3/2)$.
Thus $\partial_{p_x}=\partial_{k_x}-\partial_{k_y}/2$,
$\partial_{p_y}=(\sqrt3/2)\partial_{k_y}$, and
$A_{\rm BZ}=(2\pi)^2/(\sqrt3/2)$.
The trace excess $\operatorname{tr}g-|\Omega|\geq0$ measures departure
from the pointwise metric bound~\cite{Roy2014}.
The physical-position convention includes the orbital phases through the periodic-basis covariant derivative
$\nabla_\mu u=\partial_\mu u-ir_\mu u$, equivalently
$\partial_\mu H-i[r_\mu,H]$ before the Cartesian transformation.
Both conventions give the same Chern integral; local curvature and metric values maybe differ.

\subsection{Analytic block reductions}
\label{sec:reductions}

Invariant blocks locate contacts; eliminating loop amplitudes shows strong-loop gap-opening terms.
Write $a=t_A$, $c=\cos\phi$, $s=\sin\phi$, and
$f_\ell(j)=i^{j\ell}/2$ for $j,\ell=0,\ldots,3$ in the ordered $A$ basis.
The Fourier vectors diagonalize the loop, with eigenvalues
$a(2c,-2s,-2c,2s)$.
For the contact reductions, set $t=t_1=1$,
$F=(f_0,f_1,f_2,f_3)$, and $G=F^\dagger MM^\dagger F$.
At a momentum satisfying $M^\dagger f_2=0$, the pure-$A$ vector
has energy $E_*=-2ac$. Its six-dimensional complement has characteristic polynomial
\begin{equation}
p_6(E)=\det\!\left[E^2I_3-Ea\Lambda_{013}-G_{013}\right],
\label{eq:p6}
\end{equation}
where subscripts restrict to Fourier indices $(0,1,3)$ and
$\Lambda_{013}=\operatorname{diag}(2c,-2s,2s)$.

When $MM^\dagger f_3=2f_3$, the normalized vector
$w=M^\dagger f_3/\sqrt2$ forms a common invariant block with $f_3$:
\begin{equation}
H_2=\begin{pmatrix}2as&\sqrt2\\\sqrt2&0\end{pmatrix},\qquad
E_\pm=as\pm\sqrt{a^2s^2+2}.
\label{eq:contactblock}
\end{equation}
The five-dimensional complement has characteristic polynomial
\begin{equation}
p_5(E)=\frac1E\det\!\left[E^2I_3-Ea\Lambda_{012}-G_{012}\right],
\label{eq:p5}
\end{equation}
with $\Lambda_{012}=\operatorname{diag}(2c,-2s,-2c)$.
The determinant factor $E$ makes $p_5$ polynomial even at zero.
Matching the invariant-block and complementary energies gives, for the respective block types,
\begin{equation}
\begin{aligned}
p_6(-2ac)&=0,\\
\operatorname{Res}_E[p_5(E),E^2-2asE-2]&=0,
\end{aligned}
\label{eq:contactconditions}
\end{equation}
where $\operatorname{Res}_E$ eliminates the shared energy.
At the momenta in Table~\ref{tab:nodes}, these conditions
are polynomials of degree at most three in $v=a^2$.
Checking the complete band ordering selects central contacts among these roots.

For the strong-loop reduction, take $a>0$ and $A_\phi=aA_1$.
At fixed $\sin\phi\cos\phi\ne0$, $A_1^{-1}$ has the same four-edge
pattern, with forward entry $\zeta=1/(4\cos\phi)+i/(4\sin\phi)$
and backward entry $\zeta^*$.
Eliminating the $A$ amplitudes gives an exact Schur equation for the remaining minority-sublattice amplitudes:
\begin{equation}
E b=M^\dagger(E-aA_1)^{-1}M b.
\label{eq:schur}
\end{equation}
Set $h=-M^\dagger A_1^{-1}M$, $Z=M^\dagger A_1^{-2}M$, and
$\mathcal G=I_3+Z/a^2$.
For $E=O(a^{-1})$, expansion gives
$E\mathcal G b=(h/a)b+O(a^{-5})$.
The effective $B$ metric includes the eliminated amplitudes' norm.
Orthonormalizing with $\beta=\mathcal G^{1/2}b$ gives the Hermitian third-order matrix
\begin{equation}
H_B^{(3)}=\frac{h}{a}-\frac{\{Z,h\}}{2a^3},
\label{eq:strong_eff}
\end{equation}
since $\mathcal G^{-1/2}(h/a)\mathcal G^{-1/2}
=H_B^{(3)}+O(a^{-5})$.
This cluster expansion holds at fixed parent hoppings and nonsingular phase;
individual-band topology at leading contacts demands the third-order masses.

At a two-band contact of $h$, let $U$ be an orthonormal frame
of the degenerate subspace.
Projected velocities and the third-order perturbation have Pauli coefficients
\begin{equation}
\begin{aligned}
(v_\mu)_j&=\tfrac12\operatorname{tr}
[\sigma_j U^\dagger(\partial_\mu h)U],\\
(d_3)_j&=\tfrac12\operatorname{tr}
[\sigma_j U^\dagger(-\{Z,h\}/2)U].
\end{aligned}
\label{eq:projectedmass}
\end{equation}
The quantities $\mathcal J=|\mathbf v_x\times\mathbf v_y|$ and
$m_3=(\mathbf v_x\times\mathbf v_y)\cdot\mathbf d_3/\mathcal J$
are independent of the orthonormal frame $U$.
They determine the local gap and curvature scaling in Sec.~\ref{sec:strong}.
The cross product selects the gap-opening component of the third-order perturbation.
All derivatives here use the intracell-loop periodic $\mathbf q$ coordinates.

\section{Results}

\subsection{Parent flat bands and weak-loop stability}

\begin{figure*}[t]
	\includegraphics[width=\textwidth]{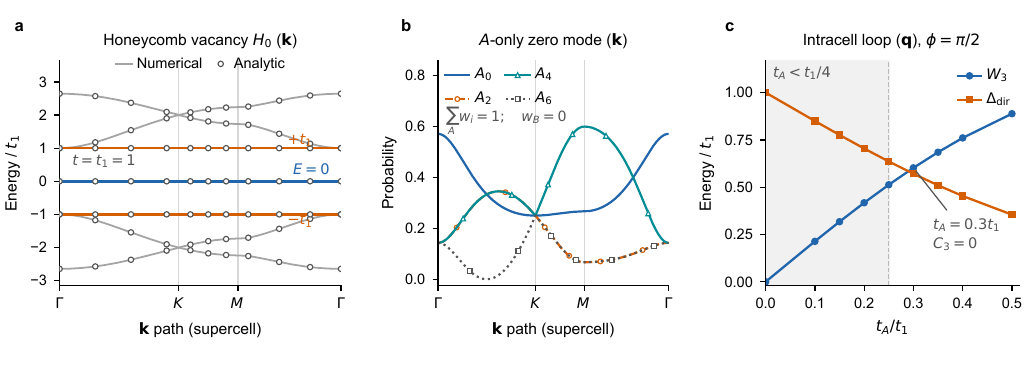}
	\caption{Algebraic flat bands and weak-loop dispersion.
		(a) Physical honeycomb-vacancy parent at $t=t_1=1$ along
		$\Gamma$--$K$--$M$--$\Gamma$, with $K=(2\pi/3,2\pi/3)$ and $M=(\pi,0)$.
		Numerical energies (lines) agree with the analytic spectrum (open circles);
		the three flat branches are $0$ and $\pm t_1$.
		(b) Sublattice-$A$ probabilities of the periodic zero mode on the same path.
		Their sum is one and the $B$ probability vanishes; some curves overlap.
		(c) Central bandwidth $w_3$ and minimum adjacent direct gap $\delta$ for the
		intracell-loop model at $\phi=\pi/2$, from a $192^2$ mesh and
		grid-seeded local refinement. The marked $t_A=0.3t_1$ point has $C_3=0$.
		Shading denotes the sufficient triviality bound $|t_A|<|t_1|/4$,
		rather than an inferred phase boundary.}
	\label{fig:parent}
\end{figure*}
The three flat branches of the parent have two different protection mechanisms.
At $t_A=0$, the bipartite form of Eq.~\eqref{eq:full} ensures
$\dim\ker M^\dagger\geq1$: the sublattice imbalance leaves a state uncoupled
to the minority sublattice. In the reordered basis, the chiral operator
$\Gamma=\operatorname{diag}(1,1,1,1,-1,-1,-1)$ satisfies
$\Gamma\widetilde H_0\Gamma=-\widetilde H_0$.
The zero mode therefore survives arbitrary intersublattice amplitudes.
The specific ring hoppings in Eq.~\eqref{eq:M} also produce two finite-energy
flat branches, whose protection must be checked separately.

To determine whether the zero mode is regular across the zone, take real
$t$ and $t_1\ne0$ and set $r=t/t_1$, $X=e^{iq_x}$, and $Y=e^{iq_y}$.
Its periodic kernel vector can be written as
\begin{equation}
u_A=\begin{pmatrix}
2\\
r(XY^2-X-X^{-1})\\
r(X^{-2}Y^{-1}-Y^{-1}-Y)\\
r(XY^{-1}-XY-X^{-1}Y^{-1})
\end{pmatrix}.
\label{eq:zero}
\end{equation}
The kernel condition $M^\dagger u_A=0$ confines this state to the majority
sublattice. Its norm is
\begin{align}
\|u_A\|^2&=4+r^2F(\mathbf q),\nonumber\\
F(\mathbf q)&=9-2[\cos2q_x+\cos2q_y\nonumber\\
&\hspace{37mm}+\cos2(q_x+q_y)],
\label{eq:F}
\end{align}
with $3\leq F\leq12$. The first component never vanishes, so
$u_A/\|u_A\|$ is a global smooth periodic section. This avoids the
obstruction to choosing an eigenvector throughout the torus and proves
$C_3=0$ for the isolated zero band. Local Berry curvature can still be
nonzero; it is its zone integral that vanishes~\cite{Xiao2010}.
At $\mathbf q=0$ and $t=t_1=1$, the normalized vector in orbital order
$(0,\ldots,6)$ is $(2,0,-1,0,-1,0,-1)^T/\sqrt7$.

The finite-energy flat pair follows from the singular values of the coupling
block. The lower three rows of $M/t_1$ have singular values $(2,1,1)$ and
a determinant of magnitude $2$. Adding the first row contributes a rank-one
positive matrix to $M^\dagger M$. One member of the degenerate pair remains
at $t_1^2$, while the other eigenvalues become
\begin{equation}
\lambda_\pm=\frac{t_1^2}{2}\left[3r^2+5\pm
\sqrt{(3r^2+5)^2-4(4+r^2F)}\right].
\label{eq:lambda}
\end{equation}
These singular values give the full parent spectrum:
\begin{equation}
\{0,\ \pm|t_1|,\ \pm\sqrt{\lambda_-},\ \pm\sqrt{\lambda_+}\}.
\label{eq:spectrum}
\end{equation}
The bound $\lambda_-\geq t_1^2$ fixes the zero band's adjacent separation
at exactly $|t_1|$. The finite-energy flat branches can touch dispersive
branches, so their existence and their isolation are different properties.
Their flatness follows from the ring amplitudes.
Figure~\ref{fig:parent}(a,b) demonstrates the physical honeycomb parent and the
momentum-dependent weights of its $A$-only zero mode.
The nonzero determinant ensures rank three everywhere on the torus,
excluding additional zero modes for $t_1\ne0$.
The lower block's twofold singular-value degeneracy is what allows a
finite-energy flat pair to survive the rank-one update.
The central branch is always separated from its neighbors under these
conditions; an individual Chern invariant for another flat branch also
demands that branch to be isolated.

To connect the Bloch zero mode to destructive interference in real space,
use the canonical representatives in Eq.~\eqref{eq:positions} and write
$x=e^{ik_x}$ and $y=e^{ik_y}$.
For the physical honeycomb parent, the zero mode is
\begin{equation}
v_A=\begin{pmatrix}
2\\
r(xy-x-1)\\
r(x^{-1}-y-1)\\
r(-xy+x-1)
\end{pmatrix},\qquad v_B=0,
\label{eq:hc_zero}
\end{equation}
It satisfies $H_{\rm hc}^{\rm per}v=0$ and has squared Bloch norm
$4+r^2\{9-2[\cos k_x+\cos k_y+\cos(k_x+k_y)]\}$.
With the Fourier convention
$v(\mathbf k)=\sum_{\mathbf n}\psi_{\mathbf n}e^{-i\mathbf k\cdot\mathbf n}$,
the finite powers in this Laurent polynomial become a finite real-space
support: ten $A$ sites in five cells for $r\ne0$, with squared norm
$4+9r^2$. The hopping-weighted amplitudes cancel at every neighboring
$B$ site, including sites outside that support. This cancellation maintains the
state compact. Its translated copies span the isolated band, although they
overlap. The construction applies to the physical parent before
same-sublattice hopping is added.

The intracell loop couples amplitudes within the majority sublattice,
allowing the zero band to disperse and breaking the parent's
$\Gamma$ anticommutation relation.
For $|t_A|\ll|t_1|$, the leading energy shift is the loop expectation value
in the isolated zero mode:
\begin{equation}
\begin{aligned}
E_3&=\frac{2t_A}{\|u_A\|^2}\operatorname{Re}
\left[e^{i\phi}\sum_{j=0}^3 u_{A,j}^*u_{A,j+1}\right]\\
&\quad+O(t_A^3/|t_1|^2),
\end{aligned}
\label{eq:perturb}
\end{equation}
Here $j+1$ is taken modulo four. The identity
$\Gamma\widetilde H(t_A)\Gamma=-\widetilde H(-t_A)$ makes $E_3$ an odd
function of $t_A$. The opposite-energy chiral partners therefore cancel
pairwise at second order, leaving the displayed cubic remainder.

Perturbation theory depicts the emerging dispersion, but spectral
isolation needs a bound valid at every momentum. The loop eigenvalues,
$2t_A\{\cos\phi,-\sin\phi,-\cos\phi,\sin\phi\}$, give the exact norm
\begin{equation}
\eta_\phi=\|A_\phi\|=
2|t_A|\max(|\cos\phi|,|\sin\phi|).
\label{eq:loopnorm}
\end{equation}
This norm bounds the displacement of every ordered energy from the
parent. With $g_0=|t_1|$, the parent has
$(E_2,E_3,E_4)=(-g_0,0,g_0)$ at all momenta, so Weyl's inequality gives
\begin{equation}
\begin{aligned}
E_2&\leq-g_0+\eta_\phi,\qquad |E_3|\leq\eta_\phi,\\
E_4&\geq g_0-\eta_\phi.
\end{aligned}
\end{equation}
Because these bounds hold across the entire zone, they constrain the
indirect gaps as well as the central bandwidth:
\begin{equation}
\min(\Delta_2,\Delta_3)\geq g_0-2\eta_\phi,
\qquad w_3\leq2\eta_\phi.
\label{eq:globalbound}
\end{equation}
The phase-independent estimate follows from $\eta_\phi\leq2|t_A|$:
\begin{equation}
\min(\delta_2,\delta_3)\geq|t_1|-4|t_A|.
\label{eq:bound}
\end{equation}
When $\eta_\phi<g_0/2$, both indirect gaps are positive and the fixed
Fermi energies $E_F=\mp g_0/2$ select three or four occupied bands.
These gaps remain open as the loop amplitude is reduced to zero, so
$C_3=C_{\mathrm{low},3}=C_{\mathrm{low},4}=0$.
The reason is that continuously scaling $t_A$ down reduces the perturbation
norm; neither fixed Fermi level can depart its gap. At the endpoint,
ordinary complex conjugation gives zero occupied-group Chern numbers for
the real-hopping parent, and Eq.~\eqref{eq:zero} supplies a global section
for its central band. Thus the deformation controls both band topology
and insulating fillings, including separations between energies at
different momenta.
The condition $|t_A|<|t_1|/4$ guarantees this result at every phase.
At $\phi=\pi/4$, the sufficient interval is wider,
$|t_A|<|t_1|/(2\sqrt2)$. These are sufficient intervals, not exact
phase boundaries. Within them, for $w_3>0$ and $\eta_\phi>0$,
the gap-to-bandwidth ratio obeys
\begin{equation}
f\geq\frac{g_0-2\eta_\phi}{2\eta_\phi}.
\label{eq:trivialflatness}
\end{equation}
Thus a large ratio is compatible with zero Chern number.

Figure~\ref{fig:parent}(c) follows how the bandwidth grows and the direct
separation shrinks along the imaginary-hopping cut.
At $t=t_1=1$, $\phi=\pi/2$, and $t_A=0.3$, the bandwidth is
$w_3=0.602t_1$ and both indirect gaps are approximately $0.509t_1$,
giving a ratio of $0.846$. The band remains isolated outside the always
weak-loop interval. To determine its topology there, we next use a
symmetry of the hopping graph that holds beyond perturbation theory.

\subsection{Graph-constrained even-Chern topology and paired band contacts}
\label{sec:symmetry}

\begin{figure*}[t]
	\includegraphics[width=0.75\textwidth]{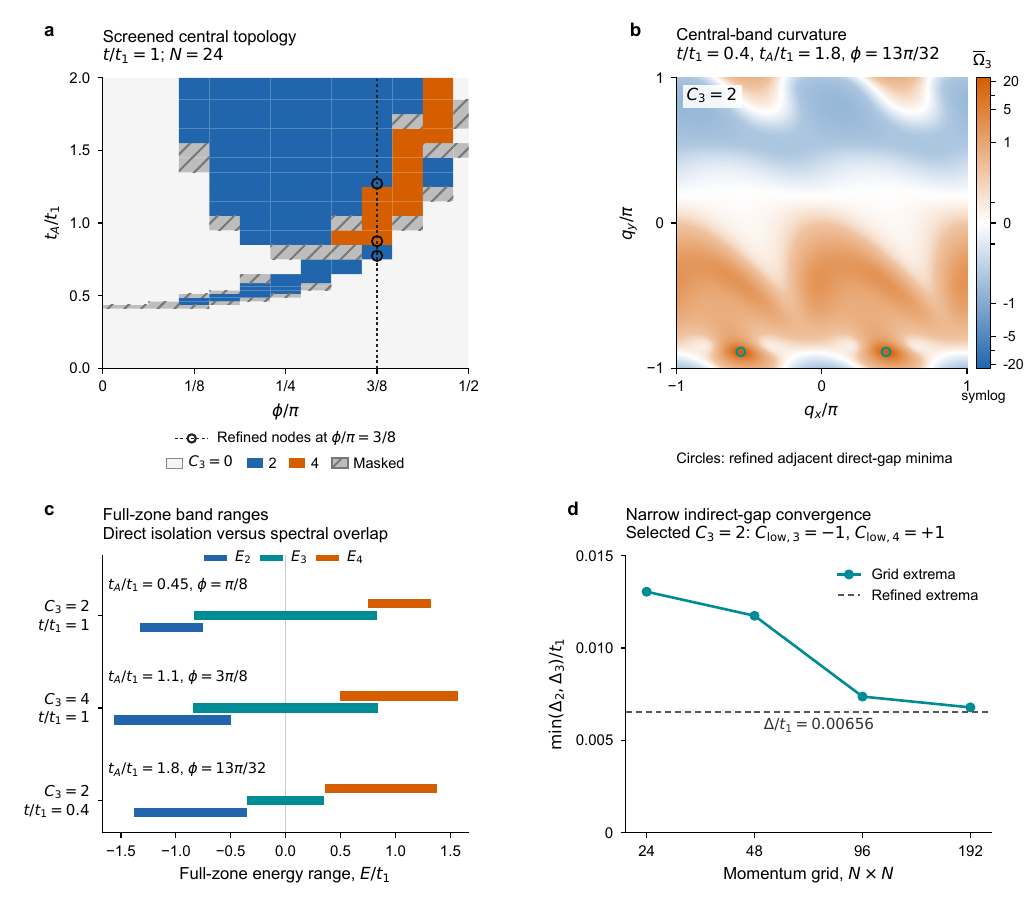}
	\caption{Even-Chern topology and spectral isolation in the intracell-loop model.
		(a) Discrete screening map at $t=t_1=1$, with accepted central-band
		$C_3=0,2,4$ values and masked unresolved samples. The three open markers on
		$\phi=3\pi/8$ are locally resolved paired-contact transitions from
		Table~\ref{tab:nodes}; the colored screening cells are not certified boundaries.
		(b) Full-zone curvature for Eq.~\eqref{eq:gappedC2}, using the periodic
		$\mathbf q$ torus. The $192^2$ overlap-plaquette integral is $C_3=2$;
		symbols mark refined lower- and upper-central-gap minima.
		(c) Refined energy ranges of $E_2,E_3,E_4$ for a directly isolated
		$C_3=2$ example, a directly isolated $C_3=4$ example, and the globally
		gapped $C_3=2$ example; their parameters are written in the panels.
		Overlap of energy ranges excludes a common insulating Fermi level.
		(d) Indirect-gap convergence for Eq.~\eqref{eq:gappedC2};
		horizontal guides are locally refined extrema estimates. Energies are in
		units of $t_1$.}
	\label{fig:literal}
\end{figure*}

\begin{table*}[]
	\caption{Nonzero central-band topology in the intracell-loop model.
		All energies are in units of $t_1=1$; $\delta$ is the smaller adjacent direct
		gap, and $\Delta$ is the smaller adjacent indirect gap. Extrema use grid-seeded local refinement.}
	\label{tab:nonzero}
	\begin{ruledtabular}
		\begin{tabular}{ccccccccc}
			$t/t_1$&$t_A/t_1$&$\phi/\pi$&$C_3$&$C_{\mathrm{low},3}$&
			$C_{\mathrm{low},4}$&$\delta/t_1$&$\Delta/t_1$&$w_3/t_1$\\
			1.00&0.450&$1/8$&2&$-1$&1&0.0494&$-0.0789$&1.66\\
			1.00&1.10&$3/8$&4&$-2$&2&0.118&$-0.347$&1.68\\
			0.400&1.80&$13/32$&2&$-1$&1&0.0641&0.00656&0.697\\
			0.375&1.60&$51/128$&2&$-1$&1&0.0212&0.00737&0.762
		\end{tabular}
	\end{ruledtabular}
\end{table*}
The loop's complex amplitudes can disperse the zero mode, but their
topological effect also depends on which orbitals and cells they connect.
Although $\Gamma A_\phi\Gamma=A_\phi$ breaks the parent's chiral
anticommutation, a spectral antisymmetry survives at every $\phi$.
In the reordered basis, define
$S=\operatorname{diag}(1,-1,1,-1,1,1,1)$ and $R=\Gamma S$.
The shift changes the parent and loop blocks in different ways:
$S\widetilde H_0(\mathbf q+(\pi,0))S=\widetilde H_0(\mathbf q)$, whereas
$SA_\phi S=-A_\phi$. Combining these identities yields
\begin{equation}
R\widetilde H(\mathbf q+(\pi,0))R=-\widetilde H(\mathbf q).
\label{eq:shiftedPH}
\end{equation}
The energy pairing is therefore
$E_n(\mathbf q+(\pi,0))=-E_{6-n}(\mathbf q)$.
For the central branch, the shifted energy has the opposite sign;
continuity then forces a zero-energy intersection. An insulating
$E_F=0$ is excluded at every phase.

The same relation constrains topology whenever the central band is
isolated. The constant unitary $R$ exchanges the lower-three- and
upper-three-band groups through an orientation-preserving translation,
so their Chern numbers are equal. Their sum with $C_3$ must vanish
because the full seven-orbital bundle is trivial. Hence
\begin{equation}
C_3=-2C_{\mathrm{low},3},\qquad
C_{\mathrm{low},4}=-C_{\mathrm{low},3}.
\label{eq:evenC}
\end{equation}
The even central-band Chern number is thus a consequence of the hopping
graph, rather than a numerical pattern in selected examples.
Only the two gaps bordering the central band are needed.
Crossings within either three-band group leave its projector well defined,
so the relation holds true even when individual outer bands have no
separate Chern invariants.

To see why the graph has this symmetry, attach parity offsets to the
orbitals in order $(0,\ldots,6)$:
\begin{equation}
\begin{aligned}
\mathbf p_0=\mathbf p_1=\mathbf p_3=\mathbf p_5&=(0,0),\\
\mathbf p_2&=(1,0),\quad\mathbf p_4=(0,1),\\
\mathbf p_6&=(1,1).
\end{aligned}
\label{eq:parityoffsets}
\end{equation}
Every parent bond satisfies $\mathbf R+\mathbf p_j-\mathbf p_i=0$
modulo two. It therefore preserves the parity of $\mathbf n+\mathbf p_i$,
splitting the parent graph into four sectors. Each added $A$ edge flips
the first parity component. The spatial sign
$P_x(\mathbf n,i)=(-1)^{n_x+p_{i,x}}$, combined with the original sublattice
sign as $\Xi(\mathbf n,i)=\Gamma_iP_x(\mathbf n,i)$, then assigns opposite
colors to the ends of every bond:
\begin{equation}
\widehat\Xi H\widehat\Xi=-H.
\label{eq:realspacechiral}
\end{equation}
This is a chiral coloring of the enlarged real-space cell.
After doubling the $x$ cell, each color has seven orbitals, so the coloring
does not create an imbalance-protected zero mode. Fourier transformation
instead gives the momentum-shift relation in Eq.~\eqref{eq:shiftedPH}.
The bond endpoints decide this constraint: independent Hermitian
amplitudes on parent bonds and independent complex amplitudes on the four
$A$ edges preserve it. Onsite terms or additional bonds need not.

An extra constraint appears at $\phi=\pi/2$.
The purely imaginary $A_\phi$ breaks ordinary conjugation
$\widetilde H(\mathbf q)^*=\widetilde H(-\mathbf q)$,
but conjugation combined with the original sublattice sign gives
\begin{equation}
\Gamma\widetilde H(\mathbf q)^*\Gamma=-\widetilde H(-\mathbf q).
\label{eq:PH}
\end{equation}
This particle-hole relation gives
$E_n(\mathbf q)=-E_{6-n}(-\mathbf q)$. At self-inverse momenta, an odd
matrix dimension requires an unpaired zero eigenvalue.
It constrains the spectrum. For the Chern integral, the useful relation
comes from integrating $A_{\pi/2}^*=-A_{\pi/2}$ with the $S$ shift identity:
\begin{equation}
\widetilde H(\mathbf q)=S\widetilde H(-\mathbf q+(\pi,0))^*S.
\label{eq:shiftedTR}
\end{equation}
Unlike particle-hole conjugation, this antiunitary relation keeps
each energy-ordered band label. It holds for arbitrary $t_A$, real
$t,t_1$, and phases $\pi/2$ or $3\pi/2$.

The same relation has a real-space interpretation:
$\widehat T=\widehat P_x\mathcal K$ with $\widehat T^2=1$.
Choose $W=1$ on sites where $P_x=+1$ and $W=i$ where $P_x=-1$.
Then $W^\dagger H W$ is real. Parent bonds preserve $P_x$; the loop bonds
join opposite spatial signs, so the transformation converts their imaginary
amplitudes into signed real hoppings. Therefore closed hopping products
have phase zero or $\pi$. Unequal imaginary loop amplitudes and real parent
hoppings obey the same argument. Its essential assumption is the intracell
graph's bond translations. Changing those translations can prevent the
staggered transformation, even when selected matrix elements look similar.
The argument therefore does not renovate time reversal for the differently
translated physical honeycomb rhombus.

For the Chern number, the torus map
$f(\mathbf q)=-\mathbf q+(\pi,0)$ preserves orientation.
Complex conjugation reverses curvature, and the constant $S$ does not
change it. The two mapped contributions cancel:
\begin{equation}
\Omega_n(\mathbf q)=-\Omega_n(f(\mathbf q)),\qquad C_n=0.
\label{eq:zeroC}
\end{equation}
For an isolated band group, its trace curvature offsets by the same
argument even if bands cross inside the group.
Thus both insulating fillings at $t_A=0.3$ have zero occupied-group
Chern number. The outer-band crossings
$(0,1)$, $(1,2)$, $(4,5)$, and $(5,6)$ prevent separate invariants for
those bands, but the isolated lower-three- and lower-four-band projectors
remain well defined.

Band isolation must also be differentiated from filling at zero energy.
In the example at $\phi=\pi/2$ and $t_A=0.3$, the negative-energy count
changes between three and four across the zone. Particle-hole symmetry
nevertheless fixes half filling at $3.5$ particles per seven-orbital cell.
This filling cuts the central band. At $t=t_1=1$, $\phi=\pi/2$,
and $t_A=0.3$, the open indirect gaps instead allow
$E_F=\mp0.5t_1$ and fillings $3/7$ or $4/7$.
Their occupied Chern numbers are both zero, so the quantized Hall
conductivity is zero.

Away from imaginary hopping, changing $\phi$ eliminates
Eq.~\eqref{eq:shiftedTR} while preserving Eq.~\eqref{eq:shiftedPH}.
Nonzero topology is then allowed, subject to the even-Chern constraint.
Figure~\ref{fig:literal}(a) screens isolated central bands in
$0<\phi<\pi/2$ and masks unresolved contacts.
Selected phases are examined on finer meshes with local gap refinement.
The colored cells remain sampled evidence; they do not determine
unsampled phase boundaries.

At $t=t_1=1$, the examples in Table~\ref{tab:nonzero} have isolated
central bands with $C_3=2$ and $4$. Direct gaps make their band invariants
well defined, but overlapping energy ranges exclude a collaborative insulating
Fermi level. Changing $t/t_1$ also opens indirect gaps on both sides at
\begin{equation}
t/t_1=0.4,\quad t_A/t_1=1.8,\quad\phi=13\pi/32.
\label{eq:gappedC2}
\end{equation}
At this point,
$(C_3,C_{\mathrm{low},3},C_{\mathrm{low},4})=(2,-1,+1)$,
consistent with Eq.~\eqref{eq:evenC}.
Fermi energies near $\mp0.352t_1$ select occupied groups with Chern
number $\mp1$. Both indirect gaps are $0.00656t_1$, and their gaps
map into one another under energy reversal.
The gap-to-bandwidth ratio is only $0.00940$: the point means
narrow-gap Chern-insulating fillings, while the central band is far from
nearly flat relative to those gaps.
Figure~\ref{fig:literal}(b--d) connects the curvature distribution to the
energy ranges and gap convergence. For both the $C_3=2$ and $4$
representatives, analytic curvature integrals agree with the overlap
invariants.

The remaining phase relations organize the equivalent and opposite-sign
examples. The $S$ transformation gives
$C_3(\phi+\pi)=C_3(\phi)$, and complex conjugation of the real parent gives
$C_3(-\phi)=-C_3(\phi)$. For an isolated band, they also imply sign reversal
under $\phi\mapsto\pi-\phi$.

\begin{table}[t]
	\caption{Paired-contact transitions at $t=t_1=1$ and $\phi=3\pi/8$.
		Listed momenta are the common-block points analyzed in the text for the
		lower $(2,3)$ contact. The upper $(3,4)$ contact is
		translated by $(\pi,0)$ modulo $2\pi$. $Q_-$ and $Q_+$ are outward
		central-band Berry charges, with orientation $(q_x,q_y,t_A)$.
		Unrounded transition amplitudes are included in the source data.}
	\label{tab:nodes}
	\begin{ruledtabular}
		\begin{tabular}{cccccc}
			$t_A/t_1$&$q_x/\pi$&$q_y/\pi$&$Q_-$&$Q_+$&$\Delta C_3$\\
			0.775&$1/3$&$-2/3$&$+1$&$+1$&$+2$\\
			0.876&$-1/6$&$1/3$&$+1$&$+1$&$+2$\\
			1.27&$-1/3$&$2/3$&$-1$&$-1$&$-2$
		\end{tabular}
	\end{ruledtabular}
\end{table}
The momentum shift in Eq.~\eqref{eq:shiftedPH} also determines how
the Chern number can change when a gap closes.
A lower-central contact at $(\mathbf q_*,E_*)$ has an upper-central
partner at $(\mathbf q_*+(\pi,0),-E_*)$.
Where the central band is isolated,
$P_3(\mathbf q+(\pi,0))=R P_3(\mathbf q)R$.
The constant unitary relation makes the two partners carry Berry flux
of the same sign. Table~\ref{tab:nodes} resolves three such pairs at
$t=t_1=1$ and $\phi=3\pi/8$.
Projecting analytic derivatives into the touching subspace gives
$h_{\rm loc}=E_*I+\mathbf d\cdot\boldsymbol\sigma$, with
$\mathbf d=J(\delta q_x,\delta q_y,\delta t_A)^T$.
The lower-contact determinants $0.265$, $0.0238$, and $-0.0722$ are
nonzero. The degeneracies therefore split linearly in independent
directions and are simple contacts in this three-parameter space.
We allocate their charge signs from the central-band flux.
Their momenta are not identified as physical honeycomb valleys:
the algebra locating a contact, its topological charge, and its
crystallographic interpretation are separate issues.

Exact common blocks illustrate why these contacts occur at fractional
momenta. At $\mathbf Q_1=(\pi/3,-2\pi/3)$ and
$\mathbf Q_3=(-\pi/3,2\pi/3)$, the condition
$M^\dagger f_2=0$ for $f_2=(1,-1,1,-1)^T/2$ leaves a pure-$A$
eigenstate with energy $-2t_A\cos\phi$ at every phase.
At $\mathbf Q_2=(-\pi/6,\pi/3)$,
$MM^\dagger f_3=2f_3$ instead gives Eq.~\eqref{eq:contactblock}.
Equation~\eqref{eq:contactconditions} then locates the fixed-momentum
branches through invariant blocks, without assuming a spatial little
group. Other positive roots still require the band-ordering test in
Sec.~\ref{sec:reductions}.

Flux through enclosing cubes gives charges
$(+1,+1)$, $(+1,+1)$, and $(-1,-1)$, accounting for
$C_3=0\to2\to4\to2$ along this cut.
The cubes enclose the contacts in $(q_x,q_y,t_A/t_1)$ space and use
outward orientation. Equal charges within each pair produce the even
changes; no second independent valley mechanism is required.
The fixed-momentum branches persist at nearby phases, but their strengths
and ordering vary. At $\phi=0.45\pi$, the central $\mathbf Q_2$ contact
occurs before $\mathbf Q_1$. These resolved branches depict selected
transitions, not a complete analytic phase diagram.

\subsection{Higher-order gap opening in the strong-loop limit}
\label{sec:strong}

Strong loop hopping separates the loop levels from a near-zero
sector, allowing an effective depiction of its narrow bands.
To control this separation, write $a=t_A>0$, $A_\phi=aA_1$, and
$\gamma_\phi=2\min(|\sin\phi|,|\cos\phi|)$.
For $\gamma_\phi>0$, the smallest loop energy magnitude is
$a\gamma_\phi$, while the maximum intersublattice norm is
$T=\sqrt{3t^2+4t_1^2}$.
If $a\gamma_\phi>2T$, Weyl bounds isolate a near-zero three-band cluster
from the negative and positive loop levels.
This cluster includes full bands $2,3,4$; it is not the lowest-three
occupied group. Turning $M$ continuously to zero preserves the external
gaps and leaves a trivial cluster, proving that its total Chern number
is zero. Virtual coupling to the loop levels gives energies of order
$a^{-1}$ and $A$-sublattice weight of order $a^{-2}$, as derived in
Sec.~\ref{sec:reductions}.
The reduction requires an invertible loop block. At
$\sin\phi\cos\phi=0$, $A_1^{-1}$ does not exist and a different reduction
is needed.

External cluster separation does not guarantee gaps inside that cluster.
At $t=t_1=1$ and $\phi=3\pi/8$, the leading matrix $h$ has four
two-band contacts at $q_y=-\pi/3$.
The lower pair lies near $q_x=3.01,-1.96$, and the upper partners are
shifted by $\pi$. Thus its central band has no isolated Chern invariant
at leading order. The coordinates displayed here are approximate;
source data retain full precision because the curvature peaks become
very narrow. The higher-order term resolves these contacts.
Equation~\eqref{eq:projectedmass} gives
$\mathcal J\simeq0.715$ and $m_3\simeq+0.167$ at both lower contacts,
with opposite masses at the upper ones.
After an $O(a^{-2})$ momentum shift, the mass opens a gap
$2|m_3|/a^3+O(a^{-5})$. It therefore predicts
\begin{equation}
\delta a^3\longrightarrow2|m_3|\simeq0.333
\qquad(t=t_1=1,\ \phi=3\pi/8).
\label{eq:strong_gap}
\end{equation}
With the convention of Eq.~\eqref{eq:QGT}, each massive continuum cone
contributes $+1/2$ to the central band. These are local flux contributions,
not separate zone invariants. Full-matrix and third-order overlap
integration both give $C_3=2$ at the tested parameters, with cluster
invariants $(-1,2,-1)$ and $C_{\mathrm{low},3}=-1$.
The cluster sum is zero, as required by the external-gap deformation.
It does not determine the occupied lower-three-group invariant: that group
contains two negative loop levels and only the lowest band of the
near-zero cluster.
Table~\ref{tab:strong} contrasts the refined gaps with the mass prediction.
The sufficient external-separation condition $a\gamma_\phi>2T$ has
threshold $2T/\gamma_\phi\simeq6.91$ on this cut.
The separation at $a=5$, below that sufficient threshold, is verified
numerically.

\begin{table}[h]
\caption{Strong-loop scaling at $t=t_1=1$ and $\phi=3\pi/8$.
$a=t_A$, with all tabulated energies in units of $t_1=1$.
Gaps use local refinement; $w_B^{\min}$ is the minimum central-band
$B$ weight sampled on a $192^2$ mesh. Chern numbers compare the full
central band with the middle band of $H_B^{(3)}$ on that mesh.}
\label{tab:strong}
\begin{ruledtabular}
\begin{tabular}{ccccc}
$a$&$\delta a^3$&$w_B^{\min}$&$C_3$&$C_B^{(3)}$\\
5&0.315&0.920&2&2\\
10&0.328&0.977&2&2\\
20&0.332&0.994&2&2\\
40&0.333&0.999&2&2
\end{tabular}
\end{ruledtabular}
\end{table}

The mass scale also controls how concentrated the Berry curvature becomes.
The local expansion predicts the peak scaling
\begin{equation}
\frac{\Omega_3(\mathbf q_{\rm gap})}{a^4}
\longrightarrow\frac{\mathcal J}{2m_3^2}\simeq12.9,
\label{eq:strong_curvature}
\end{equation}
The full matrix gives $12.9$ at $a=40$, consistent with this limit.
These quantities use the periodic intracell-loop coordinates, without a
physical honeycomb Cartesian embedding.
The velocity scale is $a^{-1}$ and the mass scale is $a^{-3}$.
Their competition narrows the curvature regions to widths
$O(a^{-2})$ while increasing the peaks to $O(a^4)$.
At $a=40$, the overlap integer is stable at $C_3=2$, yet a uniform
$384^2$ quadrature of $(2\pi)^{-1}\int\Omega_3\,d^2q$ gives only
$0.0960$: the mesh misses the sharp peaks.
Thus the integer invariant and a resolved curvature integral require
different numerical resolution checks; the latter needs node-local
refinement.
Meanwhile, $w_3=O(a^{-1})$ and the direct-gap-to-bandwidth ratio decreases
as $a^{-2}$. The indirect gaps are negative at all four tested amplitudes.
The bands become narrow in absolute energy, but their relative isolation
worsens and they do not form insulating plateaus.

\subsection{Honeycomb Chern bands, quantum geometry, and boundary states}

\begin{figure*}[t]
	\includegraphics[width=\textwidth]{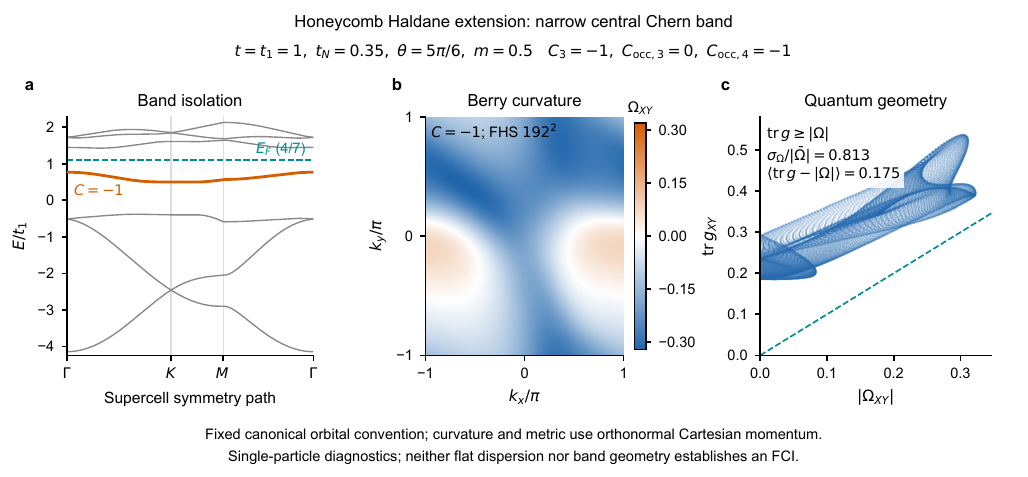}
	\caption{Narrower $C_3=-1$ band in the honeycomb NNN model,
		at $t=t_1=1$, $t_2=0.35$, $\theta=5\pi/6$, and $m=0.5$.
		(a) Physical-supercell $\Gamma$--$K$--$M$--$\Gamma$ spectrum,
		with the central band highlighted. The dashed Fermi energy in the upper gap
		selects $C_{\mathrm{low},4}=-1$.
		(b) Central-band Berry curvature in orthonormal Cartesian momentum coordinates,
		displayed over the fractional $k_x,k_y$ torus; its normalized integral
		$(2\pi)^{-1}\int\Omega\,d^2p$ is $-1$.
		(c) Cartesian metric trace plotted against the curvature magnitude over
		the zone. Points lie above the dashed trace bound
		$\operatorname{tr}g=|\Omega|$, with a nonuniform positive trace excess.
		Panels (b,c) use the same canonical periodic
		orbital convention and a $192^2$ mesh. The physical-position convention
		is compared separately in Table~\ref{tab:geometry}. Neither the flatness
		ratio nor these single-particle geometric diagnostics establish a fractional
		many-body phase. Energies are in units of $t_1$ and geometric tensors in
		units of the squared supercell-vector length.}
	\label{fig:honeycomb}
\end{figure*}

\begin{table}[h]
	\caption{Orbital-convention comparison for the central band of
		Fig.~\ref{fig:honeycomb}, evaluated on a $192^2$ mesh with orthonormal
		Cartesian derivatives. $s_{\rm opp}$ is the fraction of zone area with
		curvature sign opposite to its mean; $\mathcal T$ is the integrated trace
		metric. The Chern number is $-1$ in both conventions.}
	\label{tab:geometry}
	\begin{ruledtabular}
		\begin{tabular}{lccc}
			Convention&$\eta_\Omega$&$s_{\rm opp}$&$\mathcal T$\\
			Canonical periodic&0.813&14.8\%&2.35\\
			Physical embedding&0.537&0\%&1.40
		\end{tabular}
	\end{ruledtabular}
\end{table}

\begin{figure*}[t]
	\includegraphics[width=\textwidth]{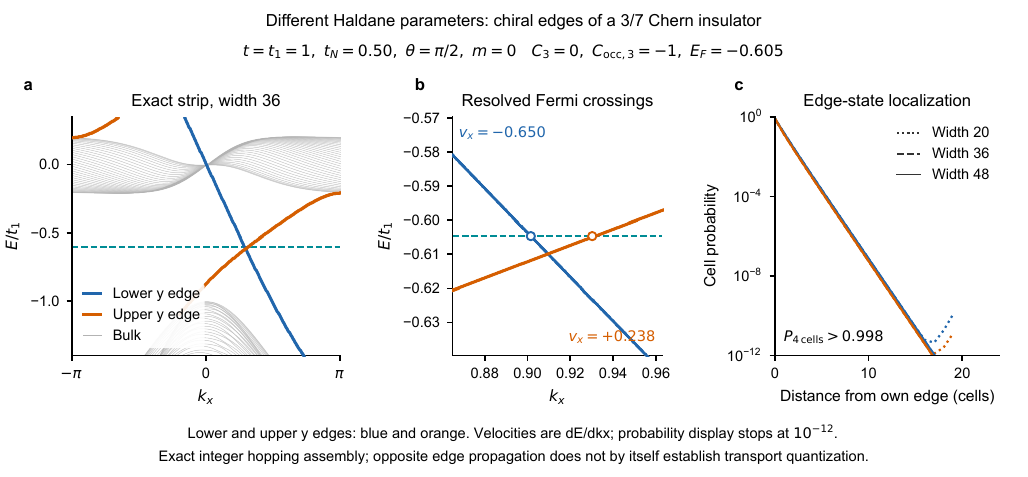}
	\caption{Clean boundary spectral flow of the honeycomb NNN model at
		$t=t_1=1$, $t_2=0.5$, $\theta=\pi/2$, and $m=0$.
		This parameter set differs from Fig.~\ref{fig:honeycomb}: the central band
		has $C_3=0$, while the occupied lower-three-band group has
		$C_{\mathrm{low},3}=-1$.
		(a) Spectrum of a strip periodic in $x$ and open in $y$, assembled directly
		from the integer real-space hoppings with 36 supercells across its width.
		Colors resolve the two edges and bulk states; the dashed Fermi level
		$E_F=-0.605t_1$ lies in the bulk lower gap.
		(b) Magnified crossings near $k_x=0.902$ and $0.930$,
		with opposite Hellmann--Feynman velocities.
		(c) Transverse cell probabilities at the two crossings for widths 20, 36, and 48,
		showing approximately $99.9\%$ weight in the respective outermost four cells.
		These are clean one-particle boundary spectra.}
	\label{fig:edges}
\end{figure*}
We now change the spatial support of the same-sublattice hopping,
rather than only its amplitude or phase.
The geometric rhombus in Eq.~\eqref{eq:physicalAA} no longer obeys the
intracell-loop shifted $R$ constraint.
Without onsite terms, imaginary same-sublattice hopping still obeys
Eq.~\eqref{eq:PH} in the periodic honeycomb gauge.
For an isolated middle band, this particle-hole relation claims
$C_3=-C_3=0$, while allowing a nonzero lower-group invariant.
Removing the intracell constraint therefore does not by itself remove
all restrictions on the central band.

To study the alternative bond pattern, we add oriented honeycomb NNN
hopping~\cite{Haldane1988}:
\begin{equation}
H_{ij}^{\rm NNN}(\mathbf k)=
t_2e^{i\nu_{ij}\theta}e^{i\mathbf k\cdot\mathbf R_{ij}},
\label{eq:NNN}
\end{equation}
Each hopping is companioned by its Hermitian reverse.
The orientation $\nu_{ij}=\pm1$ is the sign of the Cartesian cross product
of successive nearest-neighbor displacements.
We list pairs on the eight-site parent before removing orbital $7$.
The remaining pairs, including those whose enumerating paths pass through
that site, define independent physical hoppings. They do not describe
virtual propagation through the vacancy.
An optional onsite term adds $+m$ on $A$ and $-m$ on $B$.
The NNN and onsite terms substitute the rhombus perturbation.
Changing the bond support removes the intracell model's even-Chern
restriction. The purely imaginary, unstaggered limit retains the
particle-hole constraint on the middle band, but its occupied lower
group can still be topological.

The two parameter choices below test different consequences of this change.
At $t=t_1=1$, $t_2=0.5$, $\theta=\pi/2$, and $m=0$,
the lower group has $C_{\mathrm{low},3}=-1$ with indirect gap $0.791$,
while $C_3=0$. This case is used for the boundary test.
For $t_2=0.35$, $\theta=5\pi/6$, and $m=0.5$, the central band has
$C_3=-1$, bandwidth $0.273$, and indirect gaps $0.879$ and $0.657$.
Its smaller gap-to-bandwidth ratio is $2.41$.
The lower-three- and lower-four-band groups have Chern numbers
$0$ and $-1$, so the upper gap selects a nontrivial four-band filling.
Figure~\ref{fig:honeycomb} uses this second case to inspect the narrower
Chern band's dispersion and local quantum geometry.

Spectral isolation and Chern number do not describe how the wave function
changes locally with momentum. For that purpose, we compare the quantum
geometry in specified orbital conventions.
The relative curvature variation is
$\eta_\Omega=\sqrt{\langle(\Omega-\langle\Omega\rangle)^2\rangle}/|\langle\Omega\rangle|$,
and the integrated metric trace is
$\mathcal T=(2\pi)^{-1}\int_{\rm BZ}\operatorname{tr}g\,d^2p$.
The trace bound requires $\mathcal T\geq|C_3|$~\cite{Roy2014}.
Table~\ref{tab:geometry} shows nonuniform curvature and $\mathcal T>1$
in both conventions. Physical embedding reduces both diagnostics
without changing $C_3$. The integer topology is the same, although the
local geometric distribution depends on the orbital convention.

For the boundary spectrum, return to $t_2=0.5$, $\theta=\pi/2$, and $m=0$.
The integer hopping translations build a strip periodic in $x$
and open in $y$. At $E_F=-0.605$, widths of 20, 36, and 48 cells each
show one crossing per edge, with opposite velocities on opposite edges,
consistent with $C_{\mathrm{low},3}=-1$.
Figure~\ref{fig:edges} resolves these crossings and their localized states.
Localization on opposite boundaries identifies boundary spectral flow
rather than a crossing of bulk dispersions. Because the central band
is trivial at these parameters, this edge count reflects the topology
of the occupied lower group, not that of the middle band alone.

\section{Discussion}

\subsection{Graph constraints, filling, and Hall response}

The relevant design variables are the allowed bonds, their cell translations,
and their amplitudes. They influence flatness and topology in different ways.
Sublattice imbalance protects the parent zero mode, while the additional
flat branches depend on the ring amplitudes. The weak-loop bound then
shows that strong relative isolation can persist within a topologically
trivial regime. The phase information relevant to time reversal resides
in closed hopping products. At $\phi=\pi/2$, a real-space change of gauge makes the intracell hoppings
real, with closed-product phases zero or $\pi$.
The even-Chern constraint has a separate origin in the staggered chiral
coloring. It survives independent amplitude changes on the same bonds,
but its derivation depends on their cell translations. This dependence
illustrates why changing the hopping network can change the allowed topology.

The distinction between the antiunitary relations in Eqs.~\eqref{eq:PH}
and \eqref{eq:shiftedTR} clarifies the role of complex hopping.
Particle-hole conjugation relates negative and positive energies.
Shifted time reversal instead maps an energy-ordered band back to itself
and reverses its curvature. The first relation constrains the spectrum;
the second makes the Chern integral vanish. Keeping these operations
separate is essential when comparing differently translated hopping networks.

The Hall response requires a globally defined occupied subspace.
For a noninteracting insulator, its Chern number fixes
$|\sigma_{xy}|=|C_{\mathrm{occ}}|e^2/h$ for a consistent orientation
and charge convention~\cite{Thouless1982}.
Equation~\eqref{eq:shiftedPH} prevents an insulating $E_F=0$ at every
loop phase. In the half-filled imaginary-loop example, the number of
negative-energy states changes with momentum, so these states do not
constitute a separated occupied projector of fixed rank. Summing whole-band
integers cannot give an insulating Hall plateau for this occupation.
A band Chern number describes all states in that band, including those
left unoccupied by the chosen Fermi energy.

The globally gapped example in Eq.~\eqref{eq:gappedC2} makes the same
distinction at insulating fillings. Its central value $C_3=2$ describes
the change in occupied topology when that entire band is filled; it does
not assign two units of Hall conductivity to either adjacent gap.
Therefore the occupied-group invariants are the quantities to compare
with a Hall response or an edge-state count.

\subsection{Spectral isolation and quantum geometry}

The strong-loop limit illustrates why bandwidth alone can be misleading.
Along the tested cut, it decreases as $t_A^{-1}$, but the direct gaps
decrease faster, as $t_A^{-3}$. The band becomes narrower while becoming
less well isolated relative to its bandwidth.
The leading effective Hamiltonian preserves the dispersion but has band
contacts; the higher-order mass removes those contacts at finite coupling.
That mass must be kept when assigning an individual-band Chern number.
The same small gap concentrates curvature into reducing momentum regions,
so convergence of overlap integers does not ensure that a geometric
integral has resolved the curvature peaks.

The honeycomb NNN example has larger relative insulating gaps than the
tested even-Chern examples. This comparison identifies useful parameter
regimes within the present models, without selecting a global optimum or
isolating one hopping feature as the cause of the improvement.
Curvature variation and metric-trace excess characterize the wavefunctions
relevant to correlated applications~\cite{Roy2014,Liu2012}.
In specific lattice models, numerical many-body gaps have been found to
correlate with band-geometric diagnostics~\cite{Jackson2015}.
Topology places bounds on integrated geometric measures~\cite{Ozawa2021},
while leaving room for local nonuniformity.
Local geometry also depends on orbital embedding: changing orbital positions
at fixed hopping parameters can change Berry curvature without changing
the spectrum or Chern number~\cite{Simon2020}.
These geometric diagnostics need to be thought together with the
projected interaction matrix elements and many-body spectra when assessing
fractional phases~\cite{Neupert2011,Wang2011,Bergholtz2013}.

\subsection{Boundary spectra and further applications}

The boundary spectra connect the bulk occupied-group invariant to
edge-localized spectral flow~\cite{Hatsugai1993}.
To turn this correspondence into a device-conductance calculation requires
coupling the sample to leads and evaluating transmission~\cite{Datta1995,Lewenkopf2013}.
Surface Green functions can provide the lead self-energies in that
calculation~\cite{Sancho1985}, while disorder robustness must be tested
over ensembles and system sizes~\cite{Kramer1993}.
Artificial electronic, atomic, and photonic
platforms~\cite{Polini2013,Drost2017,Jotzu2014,Rechtsman2013}
provide settings for such extensions. Their connection to the present
models requires implementing the specified hopping translations and phases.

\section{Conclusions}

The seven-orbital vacancy parent provides an exact reference for tracing
the relation between flatness and topology. Its branches at $0$ and
$\pm|t_1|$ follow from sublattice imbalance and the ring structure,
with a regular periodic zero mode and a compact localized state.
For the intracell-loop extension, the hopping graph enforces even
central-band Chern numbers, paired contacts, and zero-energy intersections
at every phase. A real gauge at the imaginary-loop phase and a weak-loop
gap bound identify complementary trivial spaces.
Invariant blocks locate the $0\to2\to4\to2$ transitions on a specified
parameter cut. In the strong-loop limit, third-order masses explain why
the bands at large but finite coupling are directly isolated even though the leading
effective Hamiltonian has contacts.

Band topology and insulating topology are distinguished by the globally
gapped $C_3=2$ example, whose occupied-group invariants are $\mp1$
and whose smaller indirect gap divided by the bandwidth is below $0.01$.
Changing the bond connectivity to the honeycomb NNN extension gives
a tested $C_3=-1$ band with the same indirect-gap-to-bandwidth ratio of $2.41$.
Boundary spectra at a different parameter set resolve the topology of
the occupied group. These results supply concrete conditions for designing
isolated Chern bands from a flat-band parent, and specify the single-particle
input for further studies of interactions, disorder, and transport.

\begin{acknowledgments}
\begingroup\interlinepenalty=10000
This work is supported by Natural Science Foundation of Hebei Province
(Grant No. A2026203033) and Beijing National Laboratory for Condensed Matter
Physics (Grant No. 2024BNLCMPKF020).
\par\endgroup
\end{acknowledgments}

\bibliography{c7}
\end{document}